\documentclass[journal=apchd5,manuscript=article]{achemso}

\usepackage[version=3]{mhchem} 

\usepackage{amsmath}
\usepackage{subcaption}
\usepackage{afterpage}
\usepackage{mathtools}
\usepackage{natbib}

\DeclarePairedDelimiter\abs{\lvert}{\rvert}

\author{Tobias Vogl}
\email{Tobias.Vogl@anu.edu.au}
\affiliation{Centre for Quantum Computation and Communication Technology, Department of Quantum Science, Research School of Physics and Engineering, The Australian National University, Acton ACT 2601, Australia}
\author{Geoff Campbell}
\affiliation{Centre for Quantum Computation and Communication Technology, Department of Quantum Science, Research School of Physics and Engineering, The Australian National University, Acton ACT 2601, Australia}
\author{Ben C. Buchler}
\affiliation{Centre for Quantum Computation and Communication Technology, Department of Quantum Science, Research School of Physics and Engineering, The Australian National University, Acton ACT 2601, Australia}
\author{Yuerui Lu}
\affiliation{Research School of Engineering, The Australian National University, Acton ACT 2601, Australia}
\author{Ping Koy Lam}
\email{Ping.Lam@anu.edu.au}
\affiliation{Centre for Quantum Computation and Communication Technology, Department of Quantum Science, Research School of Physics and Engineering, The Australian National University, Acton ACT 2601, Australia}

\title[]
  {Fabrication and deterministic transfer of high quality quantum emitter in hexagonal boron nitride}


\begin{document}
\begin{abstract}
Color centers in solid state crystals have become a frequently used system for single photon generation, advancing the development of integrated photonic devices for quantum optics and quantum communication applications. In particular, defects hosted by two-dimensional (2D) hexagonal boron nitride (hBN) are a promising candidate for next-generation single photon sources, due to its chemical and thermal robustness and high brightness at room temperature. The 2D crystal lattice of hBN allows for a high extraction efficiency and easy integration into photonic circuits. Here we develop plasma etching techniques with subsequent high temperature annealing to reliably create defects. We show how different fabrication parameters influence the defect formation probability and the emitter brightness. A full optical characterization reveals the higher quality of the created quantum emitters, represented by a narrow spectrum, short excited state lifetime and high single photon purity. We also investigated the photostability on short and very long timescales. We utilize a wet chemically-assisted transfer process to reliably transfer the single photon sources onto arbitrary substrates, demonstrating the feasibility for the integration into scalable photonic quantum information processing networks.
\end{abstract}

\clearpage
Since the rediscovery of graphene\cite{Novoselov666}, the field of two-dimensional (2D) materials\cite{2053-1583-3-4-042001,10.1038/natrevmats.2017.33} has attracted great interest due to its possible applications in electronics\cite{10.1038/ncomms14948}, optoelectronics and photonics\cite{0957-4484-27-46-462001} as well as advanced sensing\cite{doi:10.1063/1.4983310} and uses in biophysics\cite{doi:10.1021/jp501711d}. More recently, the insulating 2D material hexagonal boron nitride (hBN) has drawn the attention of many researchers due to its ability to host high luminosity room temperature single photon sources (SPSs)\cite{nnano.2015.242}. In particular, the outstanding chemical and thermal stability of hBN leads to excellent robustness of the single quantum emitters, which have demonstrated long-term stable operation\cite{0022-3727-50-29-295101}. In addition, unlike NV centers in diamond, monolayered 2D material based single photon sources have almost ideal out-coupling efficiency of unity, as none of the emitters is surrounded by any high refractive index material and is not affected by Fresnel or total internal reflection\cite{PhysRevApplied.6.011001}.\\
\indent The single photon generation mechanism is based on trapping sites at point defects in the crystal lattice, which introduce energy states in the electronic band gap. While this is the generally accepted model, the exact nature of the defects remains unresolved and controversial. First principles calculations using density function theory and group theory analysis have already given some insight into the energy level structure\cite{C7NR04270A,arXiv:1709.05414}. However, the diversity of zero phonon lines (ZPLs), which vary from defect to defect, spanning the full visible spectrum\cite{doi:10.1021/acsnano.6b03602} down to the UV\cite{doi:10.1021/acs.nanolett.6b01368} show that deeper analysis and further experimental investigations are necessary.\\
\indent Single photon sources are important for quantum optics, quantum communication\cite{RevModPhys.74.145} and optical quantum computing\cite{RevModPhys.79.135}. These fields allow for the realization of unconditionally secure communication and efficient solutions for mathematically hard problems and simulations that are intractable for even the most powerful classical supercomputers. Protocols in these quantum information processing schemes require narrower emission linewidths and shorter excited state lifetimes of the trapped excitons than reported so far for single photons in hBN at room temperature. Optical quantum computing requires transform limited single photons with lifetime-bandwidth products of the order of one\cite{RevModPhys.79.135}. To date, single photons generated from 2D materials have lifetime-bandwidth products ranging from $6\times 10^3$ to $2\times 10^4$ above the transform limit at room temperature\cite{nnano.2015.242,doi:10.1021/acsnano.6b03602,doi:10.1021/acsphotonics.7b00025,doi:10.1021/acsphotonics.7b00977,PhysRevB.96.121202}.  First attempts of engineering the defect formation have been successful, using either ion irradiation\cite{doi:10.1021/acsami.6b09875}, chemical etching\cite{doi:10.1021/acs.nanolett.6b03268} or plasma etching\cite{0022-3727-50-29-295101}.\\
\indent In this letter, we describe methods to enhance the yield of particularly high quality single photon emitters in mechanically exfoliated hBN. The primary defect creation mechanism is oxygen plasma etching\cite{0022-3727-50-29-295101}, while the defect activation relies on high temperature thermal annealing\cite{nnano.2015.242}. We investigate how plasma parameters and annealing temperatures influence the formation probability and brightness of the quantum emitter and fully characterize their optical properties in terms of spectral distribution, excited state lifetime, power-dependence and photostability on short and prolonged timescales. Finally, we employ a universally applicable wet chemical transfer method for transferring the single photon sources onto arbitrary substrates, allowing for an easy integration into photonic circuits and networks.

\section*{Device fabrication}
\begin{figure*}[t!]
\centering
\begin{subfigure}{0.32\textwidth}
\centering
\includegraphics[width=0.99\linewidth,keepaspectratio]{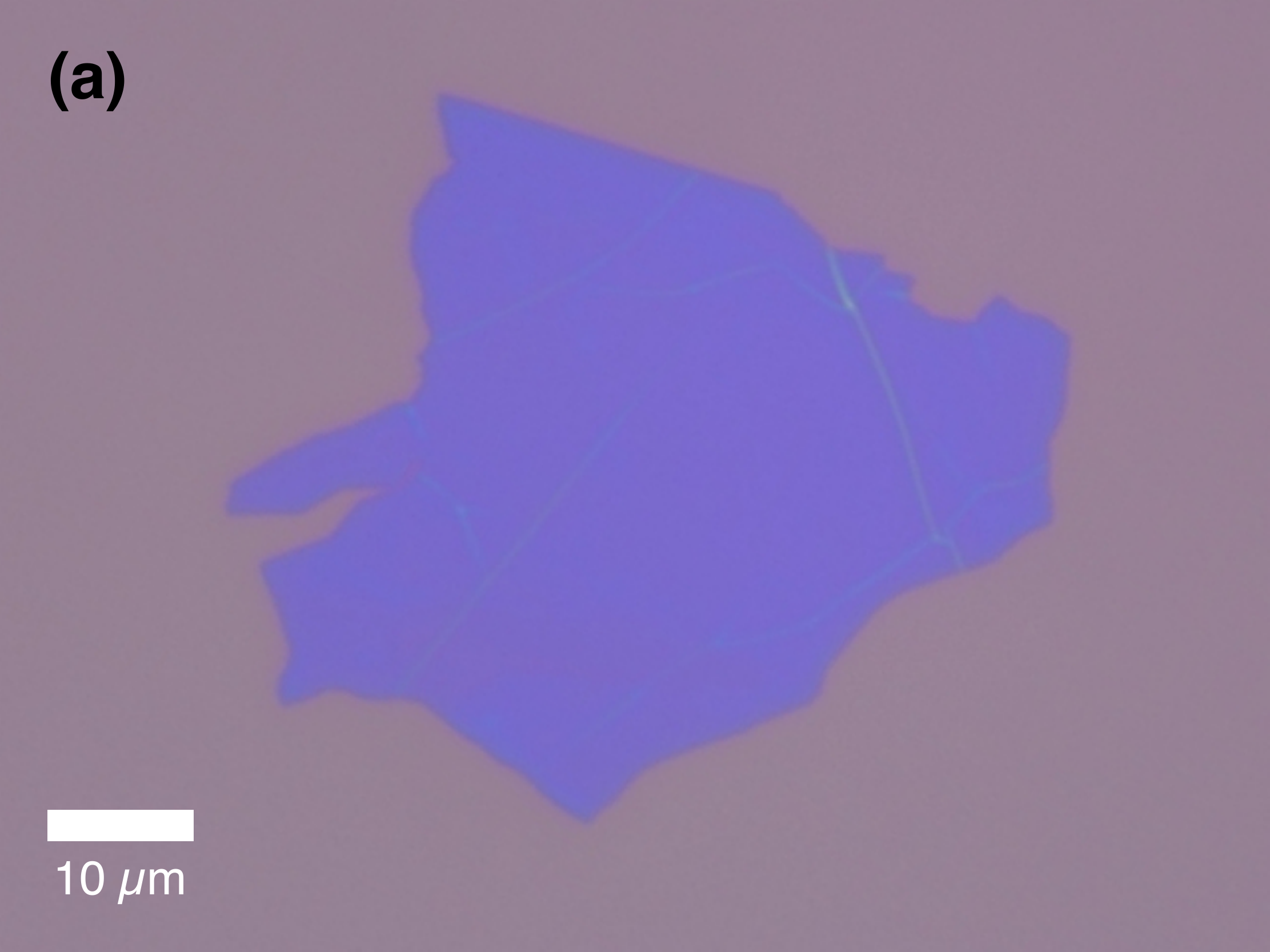}
\end{subfigure}
\begin{subfigure}{0.32\textwidth}
\centering
\includegraphics[width=0.99\linewidth,keepaspectratio]{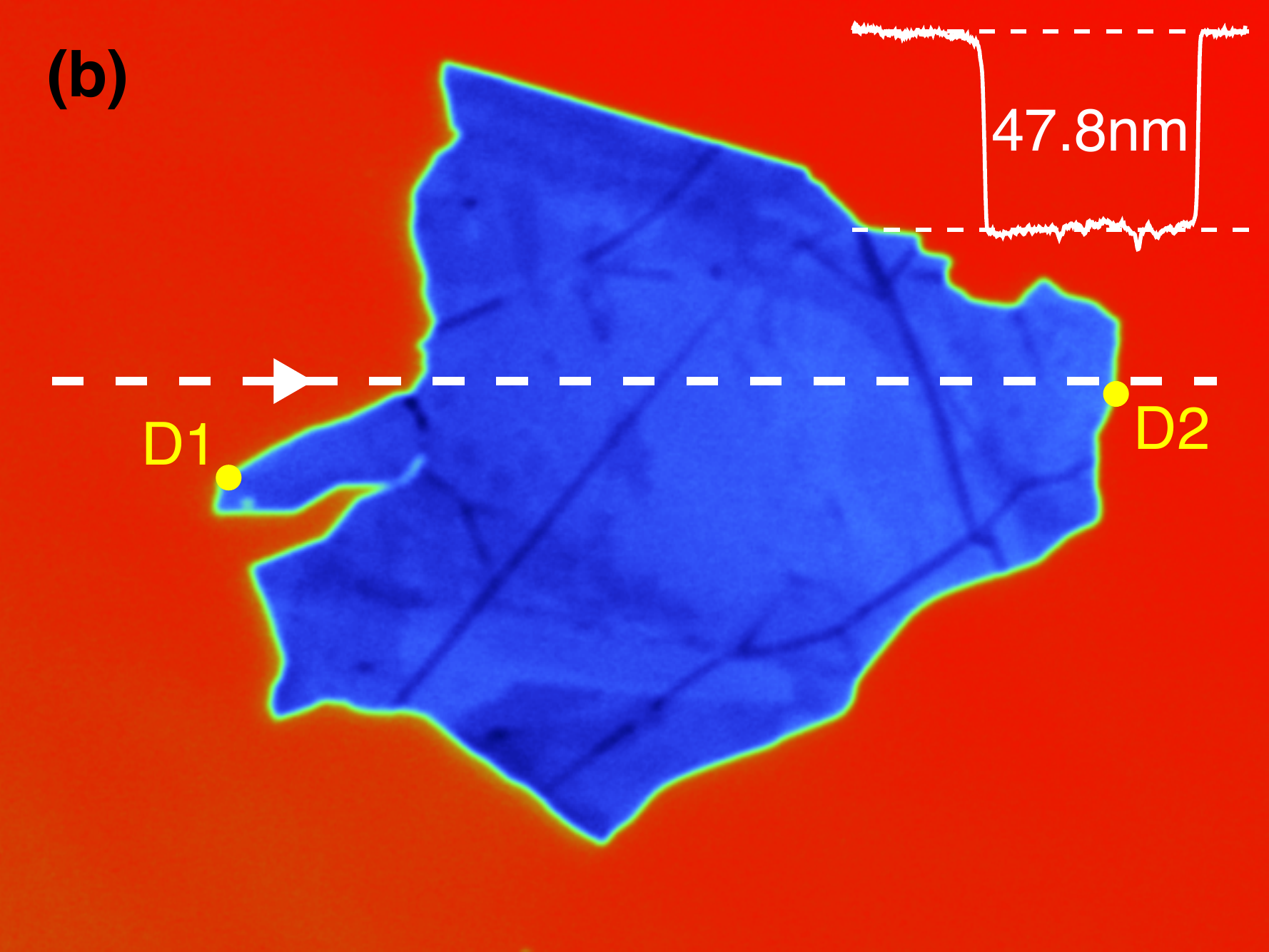}
\end{subfigure}
\begin{subfigure}{0.32\textwidth}
\centering
\includegraphics[width=0.99\linewidth,keepaspectratio]{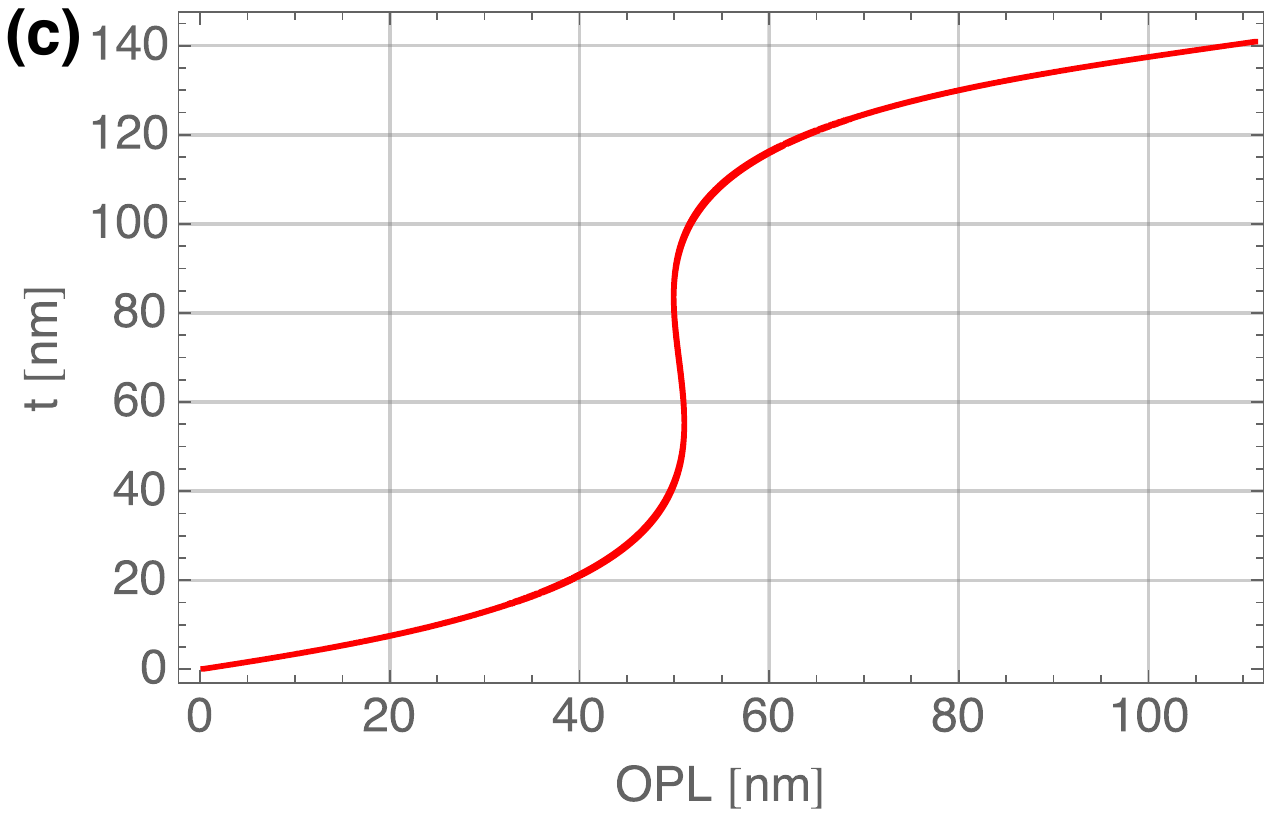}
\end{subfigure}\\
\vspace{1mm}
\begin{subfigure}{0.32\textwidth}
\centering
\includegraphics[width=0.99\linewidth,keepaspectratio]{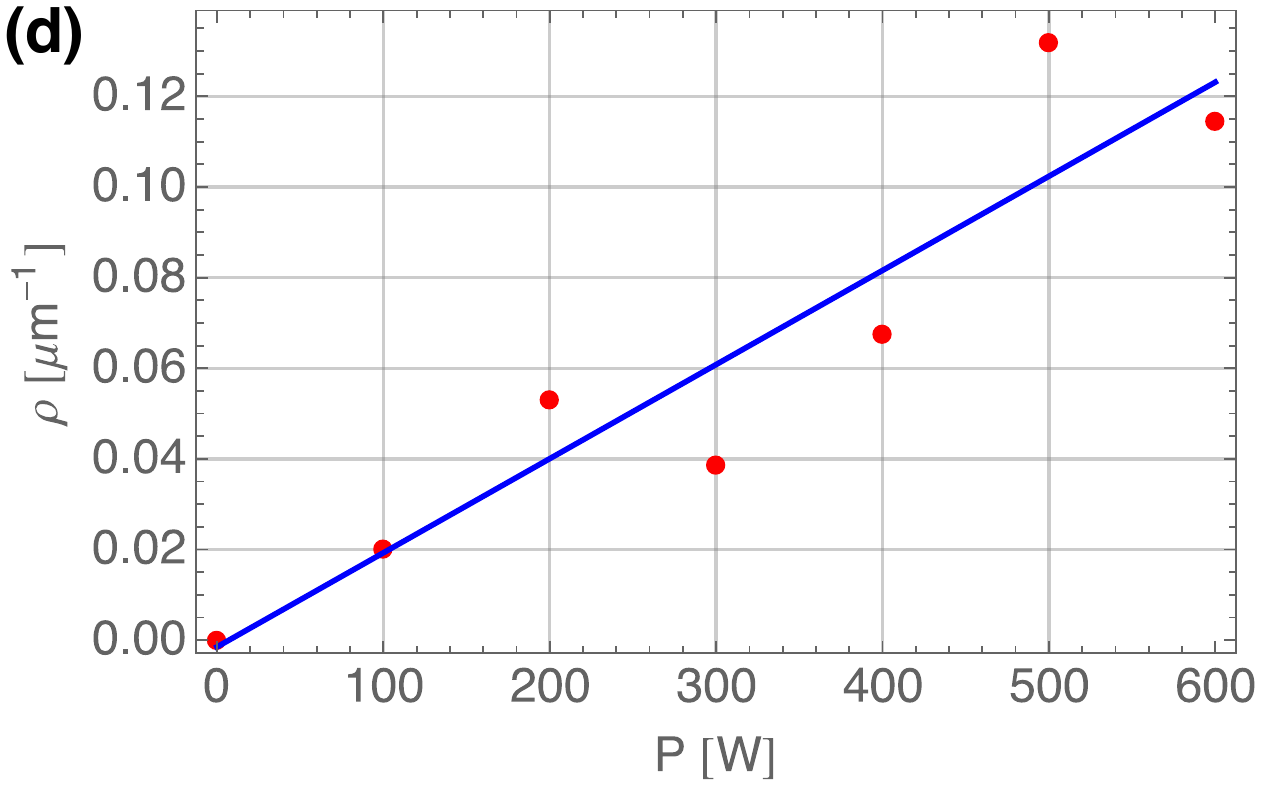}
\end{subfigure}
\begin{subfigure}{0.32\textwidth}
\centering
\includegraphics[width=0.99\linewidth,keepaspectratio]{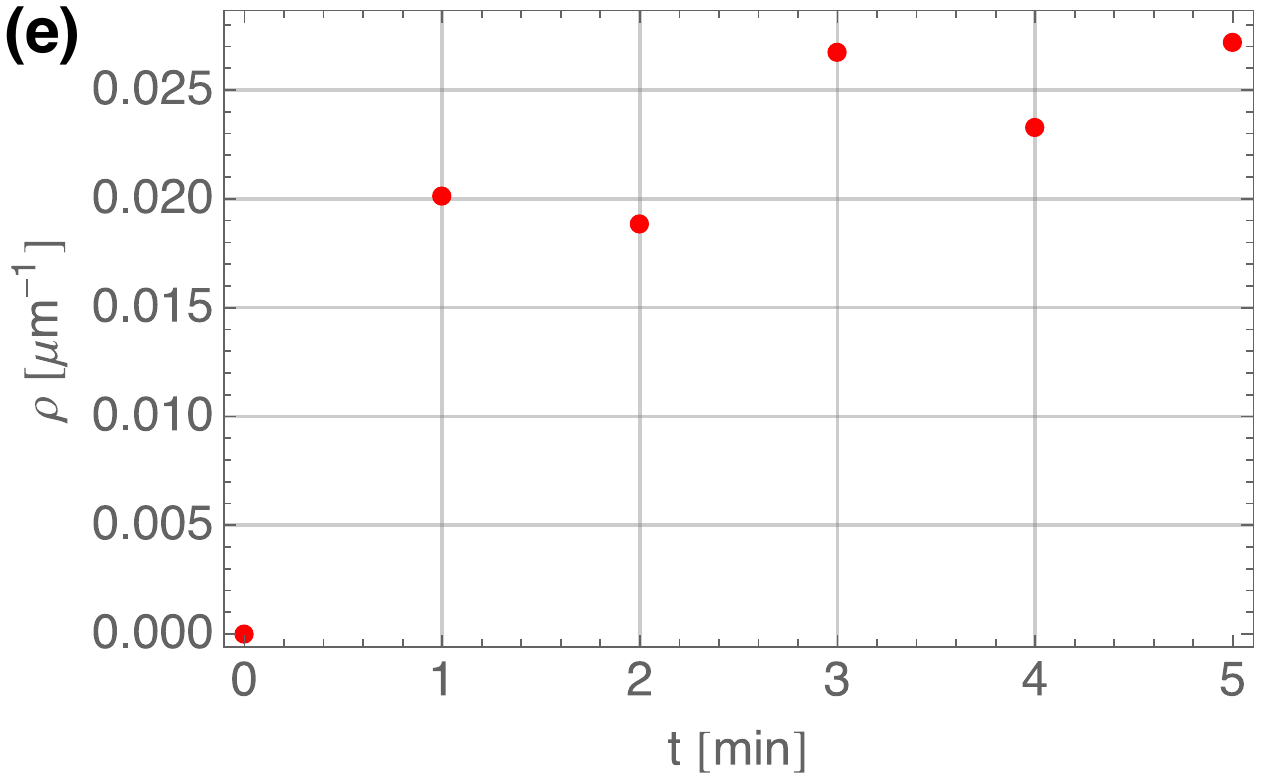}
\end{subfigure}
\begin{subfigure}{0.32\textwidth}
\centering
\includegraphics[width=0.99\linewidth,keepaspectratio]{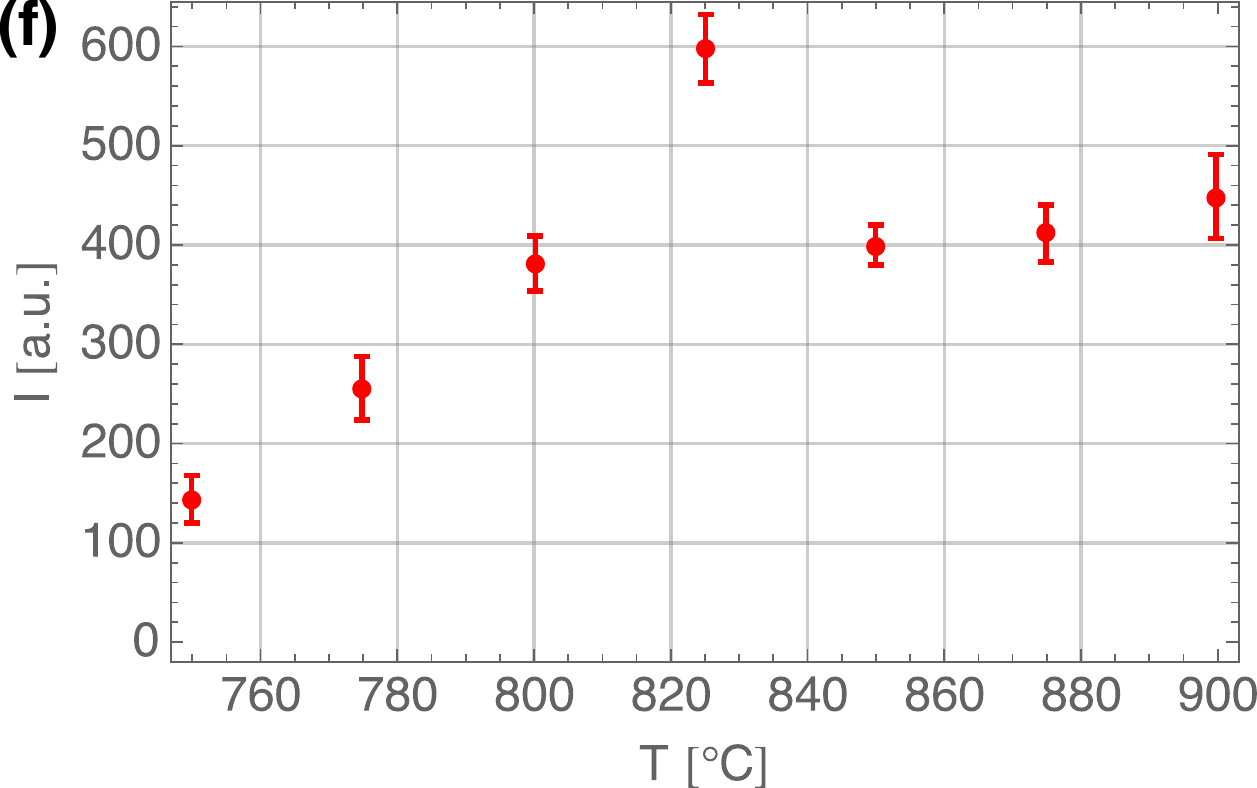}
\end{subfigure}
\caption{\textbf{Fabrication parameters. (a)} Optical microscope image of a hBN crystal. \textbf{(b)} PSI image of the crystal. The small inset shows the OPL difference along the dashed line. \textbf{(c)} RCWA simulations of the physical thickness as a function of OPL for hBN on Si/SiO$_2$, calibrated with AFM and PSI measurements. For OPLs around $50\,\text{nm}$ the simulations become ambiguous. \textbf{(d)} The linear defect density increases linearly with the plasma power. The plasma time was $1\,\text{min}$. \textbf{(e)} At a constant plasma power of $100\,\text{W}$ the linear defect density remains approximately constant for different plasma times. \textbf{(f)} Influence of the annealing temperature on the average ZPL brightness. The error bars denote the standard deviation.}
\label{fig:1}
\end{figure*}

Starting with bulk crystal hexagonal boron nitride, multi-layer flakes are mechanically exfoliated onto a viscoelastic foil. Using contrast-enhanced microscopy, thin flakes are selected by optical contrast and transferred by dry contact\cite{2053-1583-1-1-011002} to a Si substrate with a $280\,\text{nm}$ thermally grown insulating capping layer (SiO$_2$). The flake thickness is measured by phase-shift interferometry (PSI), where the optical path length (OPL) through the flake is converted to physical thickness via rigorous coupled-wave analysis (RCWA) simulations\cite{10.1038/lsa.2016.46}. By measuring the physical flake thickness of a few flakes using atomic force microscopy (AFM), the RCWA simulations of the OPL yield the refractive index of hBN of 1.849(134) for green light, which matches well previous results\cite{doi:10.1063/1.118701}. Knowing the exact refractive index allows for extrapolating the RCWA simulations, so that they serve as a conversion measure between OPL and physical thickness. A microscope image of an example flake is shown in figure $\ref{fig:1}$(a), together with the corresponding PSI map in figure $\ref{fig:1}$(b). However, for hBN on Si/SiO$_2$ the RCWA simulations yield only reliable results for (physical) flake thicknesses $<\!40\,\text{nm}$, as the simulations give an ambiguous outcome for optical path lengths $>\!50\,\text{nm}$, as shown in figure $\ref{fig:1}$(c). Thus, larger flake thicknesses are measured using AFM. We studied a large variety of flake thicknesses and found crystals with step heights ranging from $4\,\text{nm}$ to roughly $100\,\text{nm}$ were capable of hosting single photon emitter.\\
\indent In order to the create defects, the flakes are treated with an oxygen plasma and thermally annealed to activate the color centers under an argon atmosphere. In the interest of maximizing the quantum emitter yield per flake as well as optimizing the single photon spectral properties, we varied the plasma power, plasma time and annealing temperature. After plasma etching and thermal annealing the hBN samples are optically characterized in a confocal micro-photoluminescence ($\mu$PL) system scanning each flake and mapping the PL response. With the laser excitation wavelength being at $522\,\text{nm}$ ($E = 2.38\,\text{eV}$), the photon energy is well below the band gap of hexagonal boron nitride ($E_g = 5.955\,\text{eV}$\cite{10.1038/nphoton2015.77}), preventing any delocalized free excitonic emission. With the photon energy of the laser being more than a factor of 2 below bandgap and keeping the excitation power well below saturation (see next section), multi-photon excitation does not play a major role. As pure hexagonal boron nitride is optically inactive in the visible spectrum, regions with a large PL response are considered as candidates for hosting single photon emitters. During this confocal mapping, a spectrum has been taken for each scanning position. All measurements have been carried out at room temperature.\\
\indent For the sake of a fair comparison we define the average linear density of emitters per edge length $\rho = N/L$, as larger flakes are more likely to host defects, independent of the initial plasma parameters. We did not choose to take the areal density as the emitters are almost exclusively created at the boundaries of the flakes. This is a result of a low defect formation energy at the edges of the flake. We studied plasma powers varying from $100\,\text{W}$ to $600\,\text{W}$ generated from a microwave field with total plasma times ranging from $1\,\text{min}$ to $5\,\text{min}$. At this stage in the process, all samples have been subsequently annealed at $850\,^\circ\text{C}$. Figure $\ref{fig:1}$(d) shows the linear density per unit edge length as a function of plasma power, which exhibits a linear increase in defect density with plasma power (blue fit). This can be explained by the fact that at higher powers the plasma is denser, leading to the formation of more defects. When keeping the plasma power constant at $100\,\text{W}$ and varying the plasma times as shown in figure $\ref{fig:1}$(e), the linear defect density remains approximately constant. This is due to etching effects in the oxygen plasma, which is not only creating the defects, but also etching the hBN flakes layer by layer. Even though the etching rate is power-dependent, a single layer is etched faster than the timescales investigated here, so longer plasma times tend to remove already formed defects. It is worth noting, however, that the plasma field is highly anisotropic (conditioned by the gas pump, plasma generator and chamber geometry) and the plasma power varies across the plasma chamber, with the field weakening towards the center of the chamber. Thus we tried to position the substrates always at the same distance from the chamber walls, but repeating this experiment in a different plasma chamber will require using different plasma powers than the ones reported here.\\
\indent For the next part of this study we kept the plasma power constant at $100\,\text{W}$ for $1\,\text{min}$ and varied the subsequent annealing temperature from $750\,^\circ\text{C}$ to $900\,^\circ\text{C}$ under an inert Argon atmosphere. Annealing in vacuum reduces the defect yield drastically. As the defects are created during the interaction with the plasma and are only optically activated and stabilized during the annealing, we use the average brightness of the ZPL as figure of merit for a good annealing temperature, while the excitation laser power was kept constant (see figure $\ref{fig:1}$(f)). The practically usable interval of annealing temperatures spans from $800\,^\circ\text{C}$ to $850\,^\circ\text{C}$, similar to previously reported annealing temperatures\cite{nnano.2015.242,doi:10.1021/acsnano.6b03602,arXiv:1712.06938}. Lower annealing temperatures lead to weak ZPLs, where the defects are not fully optically activated, while higher annealing temperatures cause the defects to diffuse too much. Especially the latter effect is present for long annealing times as well, hence we employed rapid thermal annealing (RTA) instead of standard furnace annealing. We note that unlike in previous reports\cite{nnano.2015.242,doi:10.1021/acsnano.6b03602}, annealing in a tube furnace did not yield any bright and stable single photon emitters in our experiments.

\section*{Optical characterization}
\begin{figure*}[t!]
\centering
\begin{subfigure}{0.32\textwidth}
\centering
\includegraphics[width=0.99\linewidth,keepaspectratio]{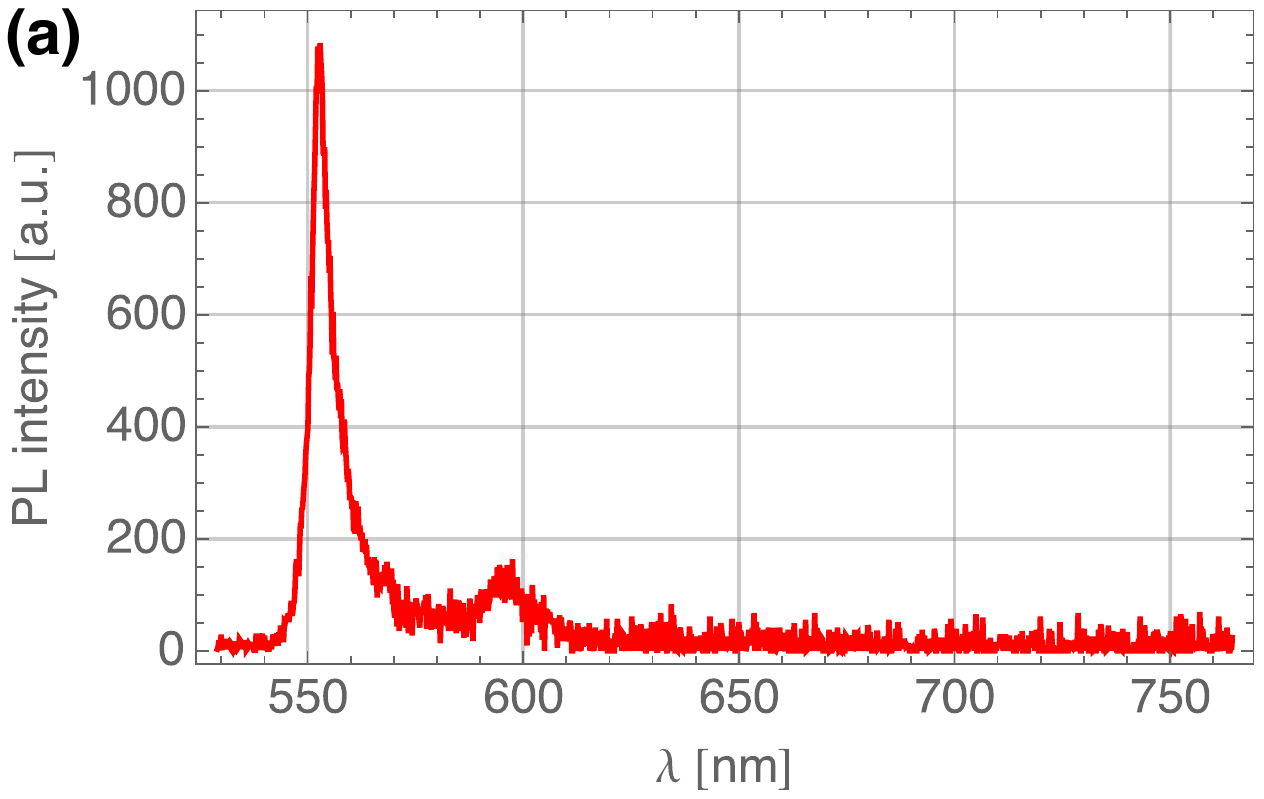}
\end{subfigure}
\begin{subfigure}{0.32\textwidth}
\centering
\includegraphics[width=0.99\linewidth,keepaspectratio]{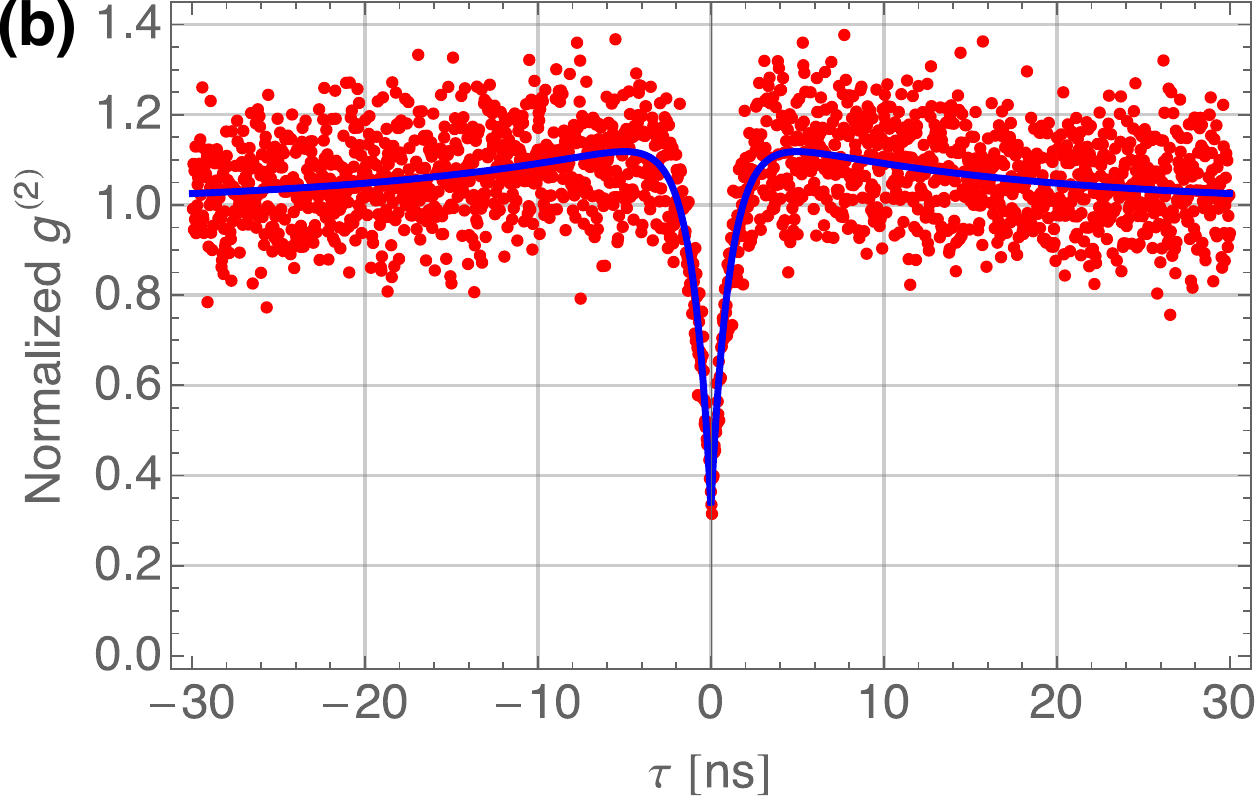}
\end{subfigure}
\begin{subfigure}{0.32\textwidth}
\centering
\includegraphics[width=0.99\linewidth,keepaspectratio]{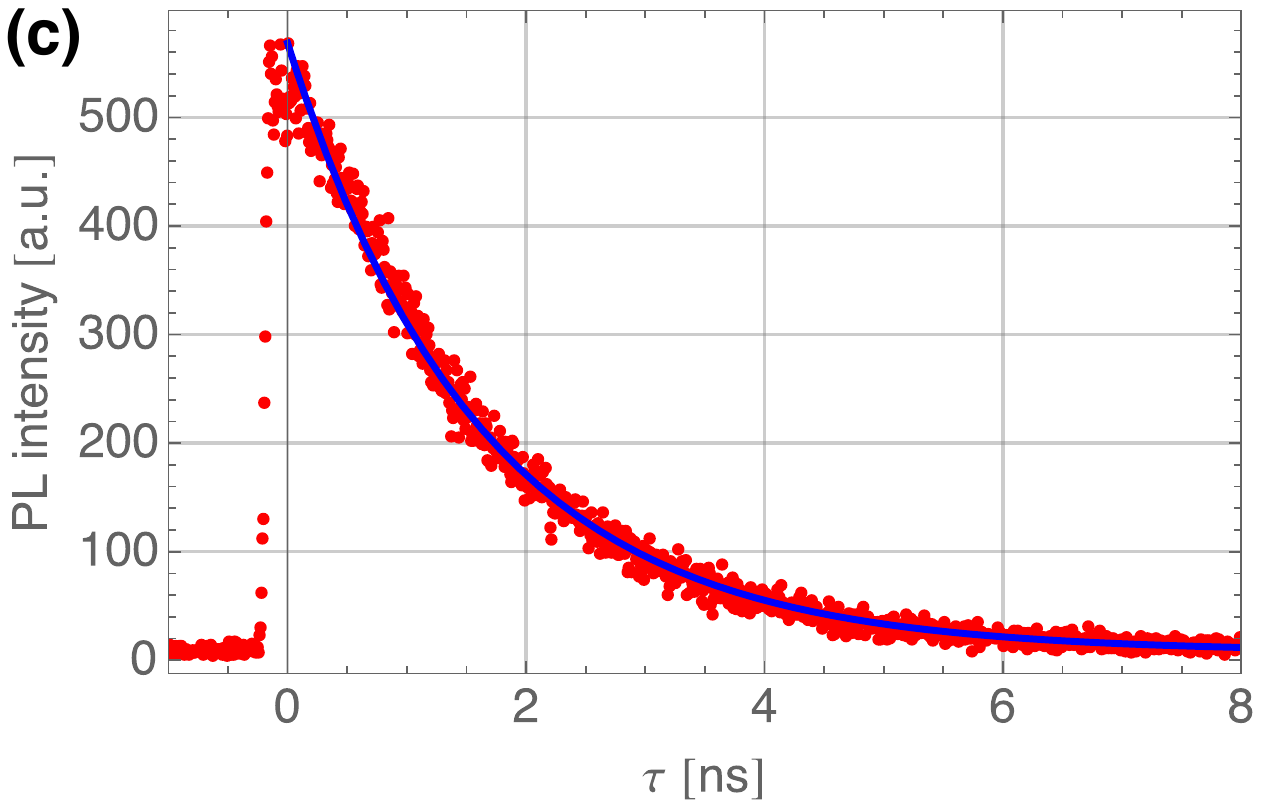}
\end{subfigure}\\
\vspace{1mm}
\begin{subfigure}{0.32\textwidth}
\centering
\includegraphics[width=0.99\linewidth,keepaspectratio]{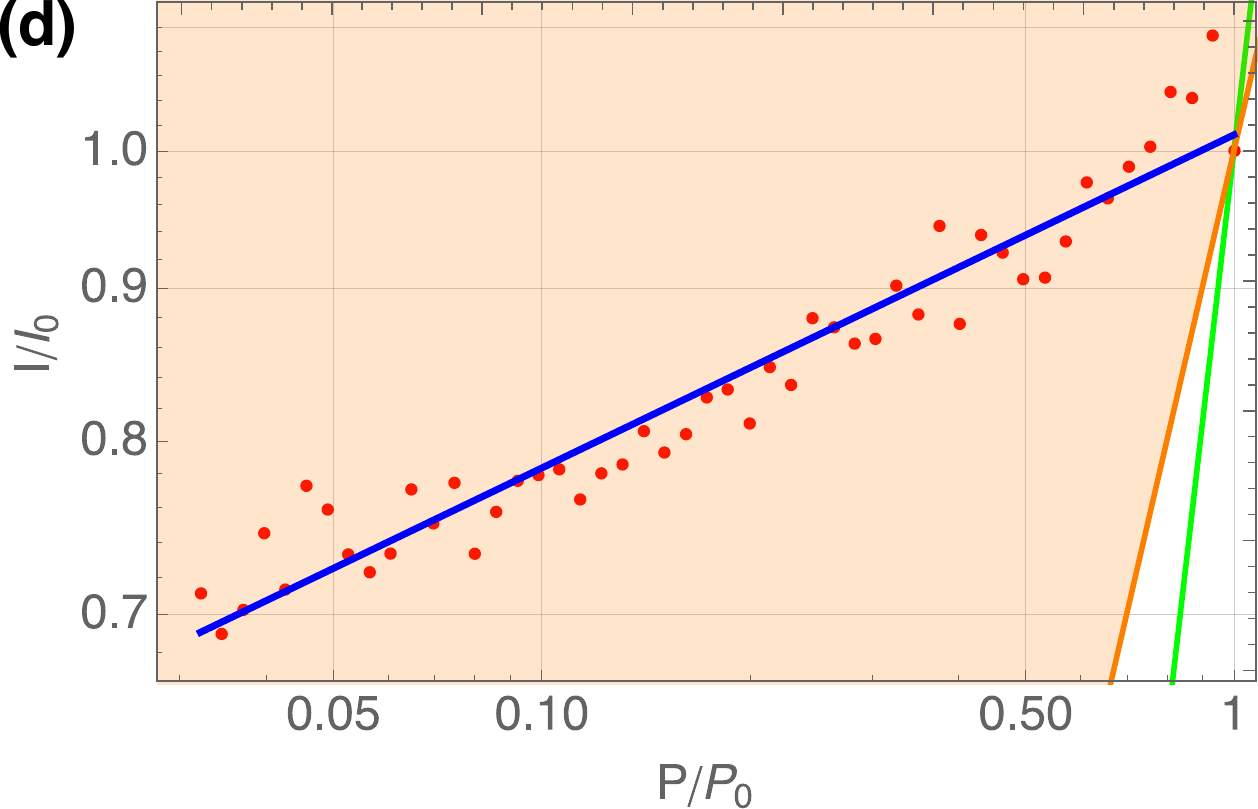}
\end{subfigure}
\begin{subfigure}{0.32\textwidth}
\centering
\includegraphics[width=0.99\linewidth,keepaspectratio]{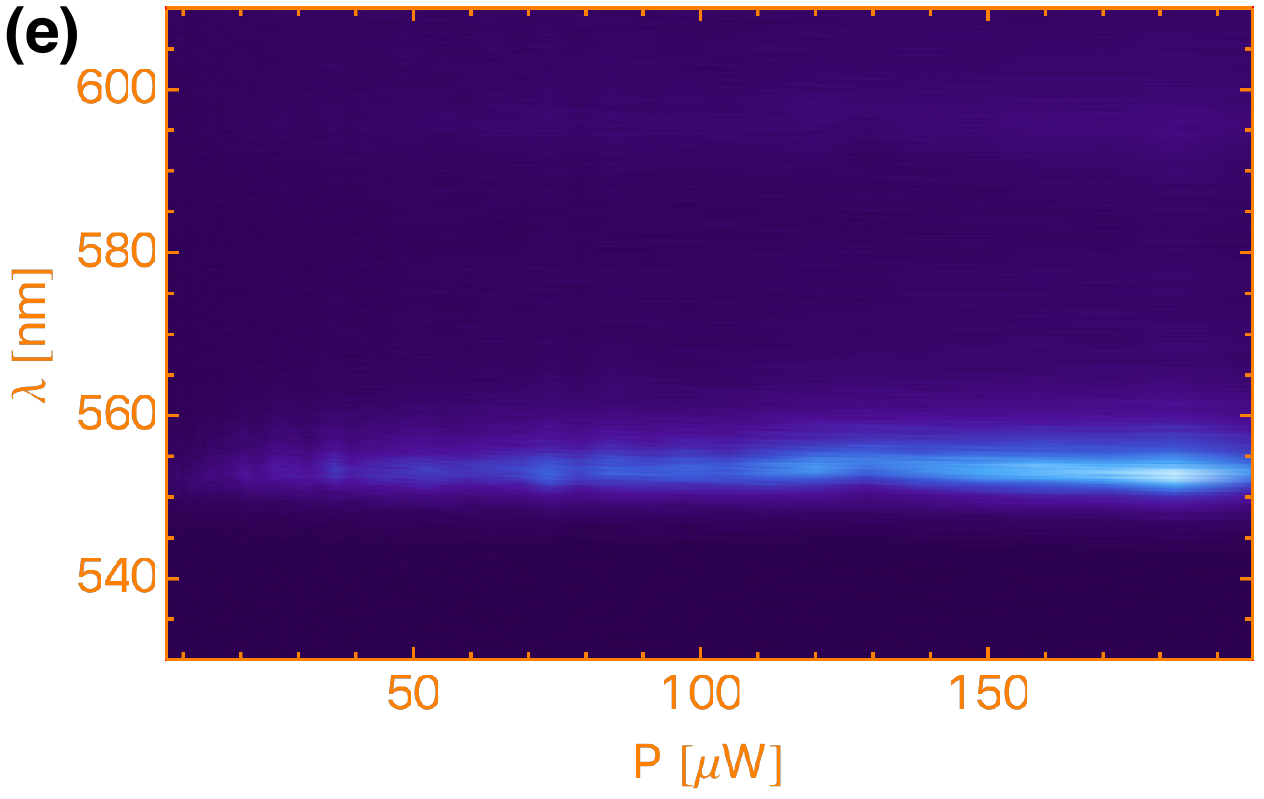}
\end{subfigure}
\begin{subfigure}{0.32\textwidth}
\centering
\includegraphics[width=0.99\linewidth,keepaspectratio]{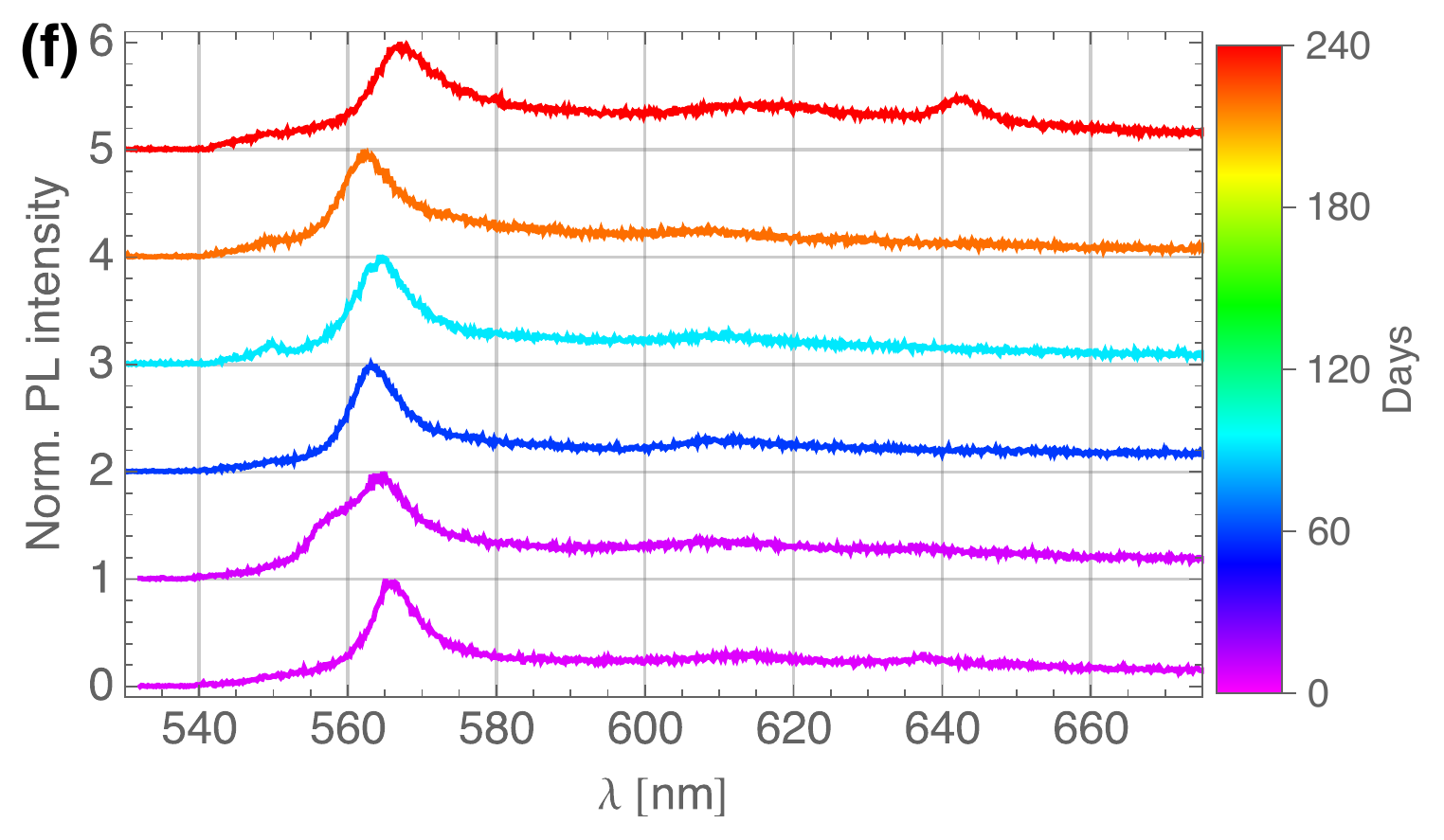}
\end{subfigure}\\
\vspace{1mm}
\begin{subfigure}{0.32\textwidth}
\centering
\includegraphics[width=0.99\linewidth,keepaspectratio]{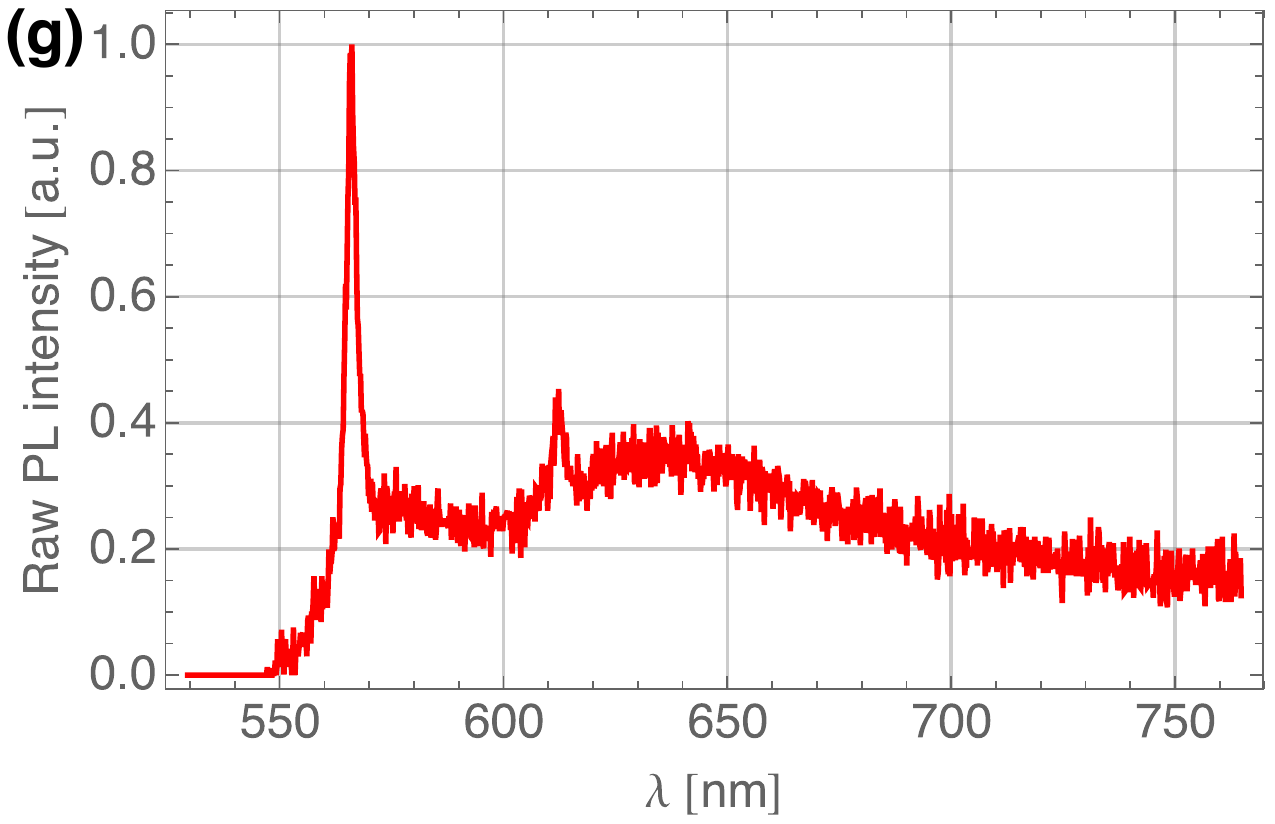}
\end{subfigure}
\begin{subfigure}{0.32\textwidth}
\centering
\includegraphics[width=0.99\linewidth,keepaspectratio]{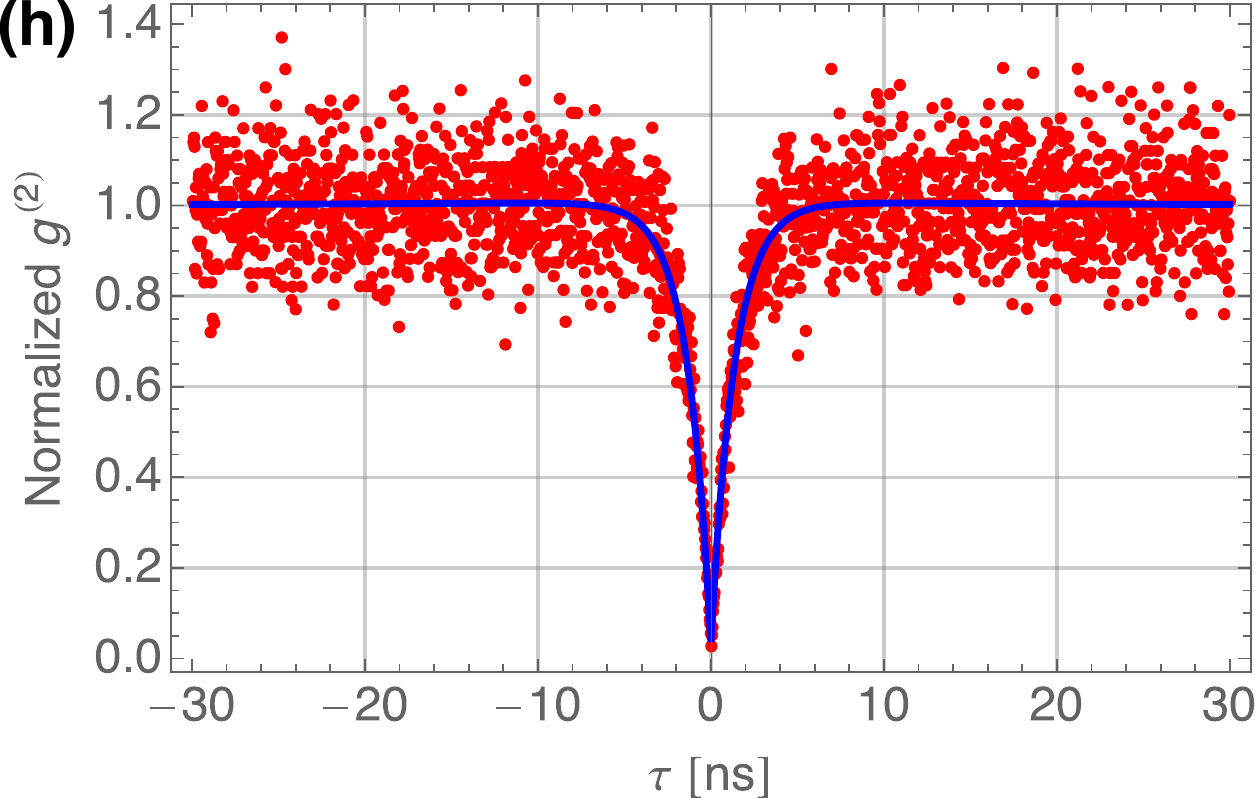}
\end{subfigure}
\caption{\textbf{Optical characterization of fabricated defects. (a)} Spectrum measured in-reflection after an ultrasteep longpass filter (opening at $530\,\text{nm}$) coupled into a high resolution spectrometer. Excited at a wavelength of $522\,\text{nm}$, the ZPL is at $553.23(5)\,\text{nm}$ with a linewidth of $2.82(10)\,\text{nm}$. \textbf{(b)} Second-order correlation function dipping at zero time delay to $0.330(28)$ (obtained from fit). \textbf{(c)} Time-resolved photoluminescence using an ultrashort pulsed laser, revealing an excited state lifetime of $\tau=1.123(7)\,\text{ns}$. \textbf{(d)} Log-Log-plot of the photoluminescence response as a function of excitation power. The orange-shaded area (slope $\alpha < 1$) indicates emission from defects, while the orange line ($\alpha=1$) corresponds to free excitonic emission and the green line ($\alpha=2$) bi-excitonic emission. The slope $\alpha = 0.350(54) < 1$ of the linear fit confirms defect emission. \textbf{(e)} Spectrally resolved power-dependence measurement. The emitter showed some power-dependent photobleaching. \textbf{(f)} Long-term stability of a defect over a duration of 8 months (normalized and vertically offset for clarity). The center of the ZPL remains stable within $\pm 2.5\,\text{nm}$, while its linewidth increases with time. \textbf{(g)} Spectrum of the best single photon emitter we found with a ZPL at $566.04(4)\,\text{nm}$ and a linewidth of $1.31(7)\,\text{nm}$. $8.7\,\%$ of the emission is into the ZPL. \textbf{(h)} The second order correlation of the defect with the spectrum shown in (g) dips to $0.033(47)$ at zero time delay.}
\label{fig:2}
\end{figure*}

We now turn to a full characterization of the single photon emitters. The sample crystal shown in figure $\ref{fig:1}$(a) hosts two defects, with their positions labeled D1-D2 in figure $\ref{fig:1}$(b). The spectrum of D1 is shown in figure $\ref{fig:2}$(a) and from a fit we extract the ZPL at a wavelength of $553.23\left(5\right)\,\text{nm}$ and a linewidth of $2.82\left(10\right)\,\text{nm}$. Unless stated else, we use a $95\,\%$ confidence interval for the uncertainties, calculated using Monte Carlo simulation methods. This defect emitted $82.4\,\%$ into its ZPL. Using a Hanbury Brown and Twiss (HBT) type interferometer, we measure the single photon purity or second-order correlation function (see figure $\ref{fig:2}$(b)), with $g^{(2)}\left(\tau=0\right)$ dipping to $0.330(28)$, obtained by fitting a three-level system with ground and excited states as well as a meta-stable shelving state: 
\begin{align}
g^{\left(2\right)}\left(\tau\right)=1-Ae^{-\abs{\tau-\mu}/t_1}+Be^{-\abs{\tau-\mu}/t_2}
\end{align}
where $t_1$ and $t_2$ are the excited and meta-stable state lifetimes respectively, $\mu$ accounts for different electrical and optical path lengths in the HBT interferometer and $A$ and $B$ are the anti-bunching and bunching amplitudes. The experimental data has been normalized such that for very long time delays $g^{\left(2\right)}\left(\tau\rightarrow \infty\right)=1$. In all correlation function measurements no background correction has been applied, as these measurements are not yet dark count limited, the dark counts from the used detectors are very low compared to the single photon count rate (see methods). The $\mu$PL system is equipped with an ultrashort pulsed laser with $300\,\text{fs}$ pulse length at a repetition rate of $20.8\,\text{MHz}$, allowing us to measure the exciton lifetime as well, with the lifetime of D1 shown in figure $\ref{fig:2}$(c). A fit of a single exponential decay reveals an excited state lifetime $\tau$ of $1.123(7)\,\text{ns}$. The shelving state lifetime is not accessible from this measurement due to its weak transition and longer lifetime. However, the decay time is consistent with the correlation function measurements. Together with its linewidth $\Delta\nu=\frac{c\Delta\lambda}{\lambda^2}$ this yields a lifetime-bandwidth product of 3102, still far above the transform limit, but better than any reported emitter in hBN so far. In addition we measured the PL intensity as a function of excitation power, which is described by
\begin{align}
I\left(P\right)=\frac{I_{sat}\cdot P}{P+P_{sat}} + I_d
\end{align}
with $I_{sat}$ and $P_{sat}$ being the saturation intensity and power respectively and $I_d$ is the dark count intensity. From a fit we extract $P_{sat}=142.6(685)\,\mu\text{W}$. Together with a focal spot diameter of $0.67\,\mu\text{m}$ and a duty cycle of $6.24\cdot 10^{-6}$ this amounts to a peak intensity of $1.62\,\text{GW cm}^{-2}$, which is still below damage threshold. Figure $\ref{fig:2}$(d) shows this measurement (red dots) together with the fit (blue line) on a Log-Log-scale, which confirms the defect nature of the emission: A slope of $\alpha=1$ indicates free excitonic emission (orange line), while $\alpha=2$ reveals the presence of bi-excitons (green line) and $\alpha<1$ verifies trapped excitons from a defect (orange-shaded area)\cite{10.1038/nphoton.2013.179}. The power-dependence of D1 has a slope of $\alpha=0.350(54)$, clearly in the defect emitter region. Furthermore, we measured the power-dependent photostability, which is shown in figure $\ref{fig:2}$(e). Defect D1 showed some power- as well as time-dependent photobleaching. The power-dependent photobleaching causes the deviations from the linear fit in figure $\ref{fig:2}$(d) and is also reason for the large confidence interval on $P_{sat}$. However, we have also found emitters that were photostable. Finally, we also look at the long-term stability, meaning repeating all measurements above for a subset of samples over a time span of more than 8 months. In between measurements the samples were stored under normal atmosphere in air. Figure $\ref{fig:2}$(f) shows the spectra for different days, all normalized and offset vertically for clarity. With the center of the ZPL being constant within $\pm 2.5\,\text{nm}$, its linewidth increases over time from $4.38(13)\,\text{nm}$ to $6.61(25)\,\text{nm}$. Other optical properties such as $\tau$, $\alpha$ and $g^{\left(2\right)}\left(0\right)$ are varying as well, without showing a clear trend in the case of $\alpha$ and $g^{\left(2\right)}\left(0\right)$ (see supplementary information S1), while $\tau$ shortens with an increase in linewidth (see also the next section). The variations can be explained by the fact that 2D materials typically oxidize in an ambient environment. The stability in air is ultimately controlled by the oxygen dissociative absorption barrier and is also affected by defects present. As the host crystal is very thin and the interactions within the crystal are strong, already small variations can cause large changes in the photophysical properties. Isolating the crystal from any coupling to the environment, such as through encapsulation, can improve the long-term stability, but the influence of the encapsulation layer must be investigated. Nevertheless, the defects maintain their single photon emission properties on short and very long timescales and at the same time keep the photophysical properties (for a 2D material) constant within the reported limits.

\section*{Correlating optical properties}
The optical properties as described in the previous section are by no means representative for all defects, but are typical photophysical properties. The optical properties in terms of spectral distribution, excited state lifetime, power-dependence, photostability and second-order correlation function vary not only from flake to flake, but also from defect to defect hosted by the same flake. As reported previously\cite{doi:10.1021/acsnano.6b03602}, the ZPLs cover the full visible spectrum below the excitation photon energy. In our experiments the quantum emitter ZPLs span a range from $550\,\text{nm}$ to $720\,\text{nm}$, with the lower limit set by a longpass filter used to filter out the excitation laser and the upper limit set by the spectrometer bandwidth. The linewidths vary from as low as $1.31(7)\,\text{nm}$ (see figure $\ref{fig:2}$(g)) to $11.6(4)\,\text{nm}$ at room temperature, while the exciton lifetimes span a smaller range from $294(3)\,\text{ps}$ to $1.32(1)\,\text{ns}$\footnote{$1.31\,\text{nm}$ linewidth and $1.32\,\text{ns}$ lifetime are not from the same defect, the same holds for $11.6\,\text{nm}$ and $294\,\text{ps}$.}. This is more than one order of magnitude faster than any previously reported excited state lifetime in hBN, in fact, all of the defects have shorter lifetimes than the fastest previously reported ones (see also supplementary information S3). The single photon purities characterized by $g^{\left(2\right)}\left(0\right)$ vary from $0.033(47)$ to $0.480(38)$ (excluding any emitter with $g^{\left(2\right)}\left(0\right)>0.5$, which are considered ensembles). A single photon purity with $g^{\left(2\right)}\left(0\right)=0.033(47)$ (see figure $\ref{fig:2}$(h)) in hBN is only matched by emitters coupled to plasmonic nanocavity arrays\cite{doi:10.1021/acs.nanolett.7b00444}, with $g^{\left(2\right)}\left(0\right)=0.02-0.04$. This defect has a time-bandwidth product of 1389. The slopes of the power saturation vary from $0.290(56)$ to $0.942(43)$ for different defects. Across all defects, the optical properties are randomly distributed with the exception of the zero phonon line, which with a $53\,\%$ chance is between $550\,\text{nm}$ and $570\,\text{nm}$ (see supplementary information S3).\\
\indent The natural question then arises whether there is any correlation between the optical properties and especially between the optical properties and the fabrication parameters or geometrical features of the host crystal flakes. By studying a large variety of flakes and cross-correlating optical properties, we found that a narrow linewidth correlates with a longer excited state lifetime, even though the single photons are still above the transform limit (see supplementary information S2). The smallest time-bandwidth product was 807 at room temperature, which is one order of magnitude smaller than any previously reported value. The mean of 3782 for this product shows the higher quality of the emitters, compared to emitters fabricated by other methods\cite{nnano.2015.242,doi:10.1021/acsnano.6b03602,doi:10.1021/acsphotonics.7b00025,doi:10.1021/acsphotonics.7b00977,PhysRevB.96.121202}.. So far it seems that neither the fabrication parameters, nor the physical crystal thickness at the defect position have any influence on the emission spectra, lifetimes, purities or $\alpha$. The latter demonstrates that the interaction of the in-plane dipole with surrounding layers is probably small. Even though it remains obscure why defects formed by plasma etching of the host crystals perform better and have particularly short lifetimes, we have seen that this process creates reliably a large number of higher quality single photon emitter (see also supplementary information S3). In total we studied more than 300 flakes hosting more than 200 defects. Each flake hosted between 0 and 7 defects, with the average number being 2.55 (not counting the flakes hosting no defect).

\section*{Deterministic transfer of quantum emitters}
\begin{figure*}[t!]
\centering
\begin{subfigure}{0.3\textwidth}
\centering
\includegraphics[width=0.99\linewidth,keepaspectratio]{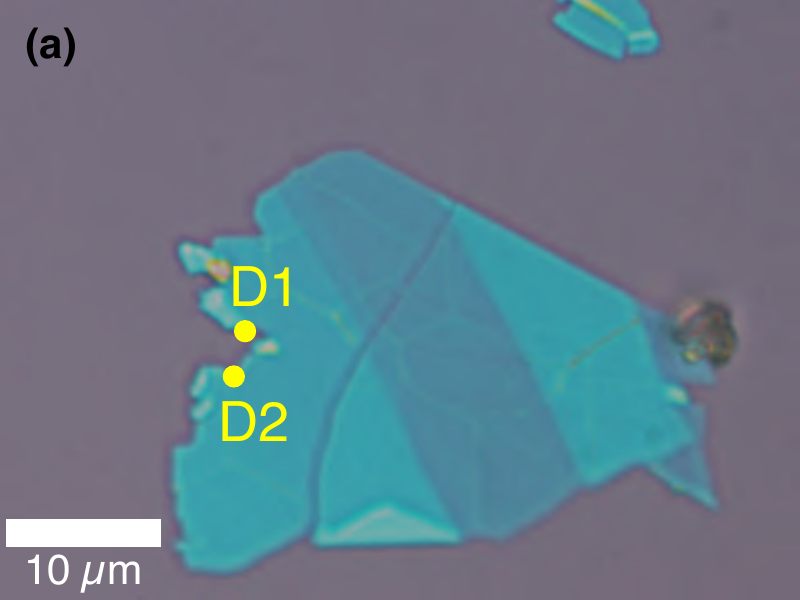}
\end{subfigure}
\begin{subfigure}{0.34\textwidth}
\centering
\includegraphics[width=0.99\linewidth,keepaspectratio]{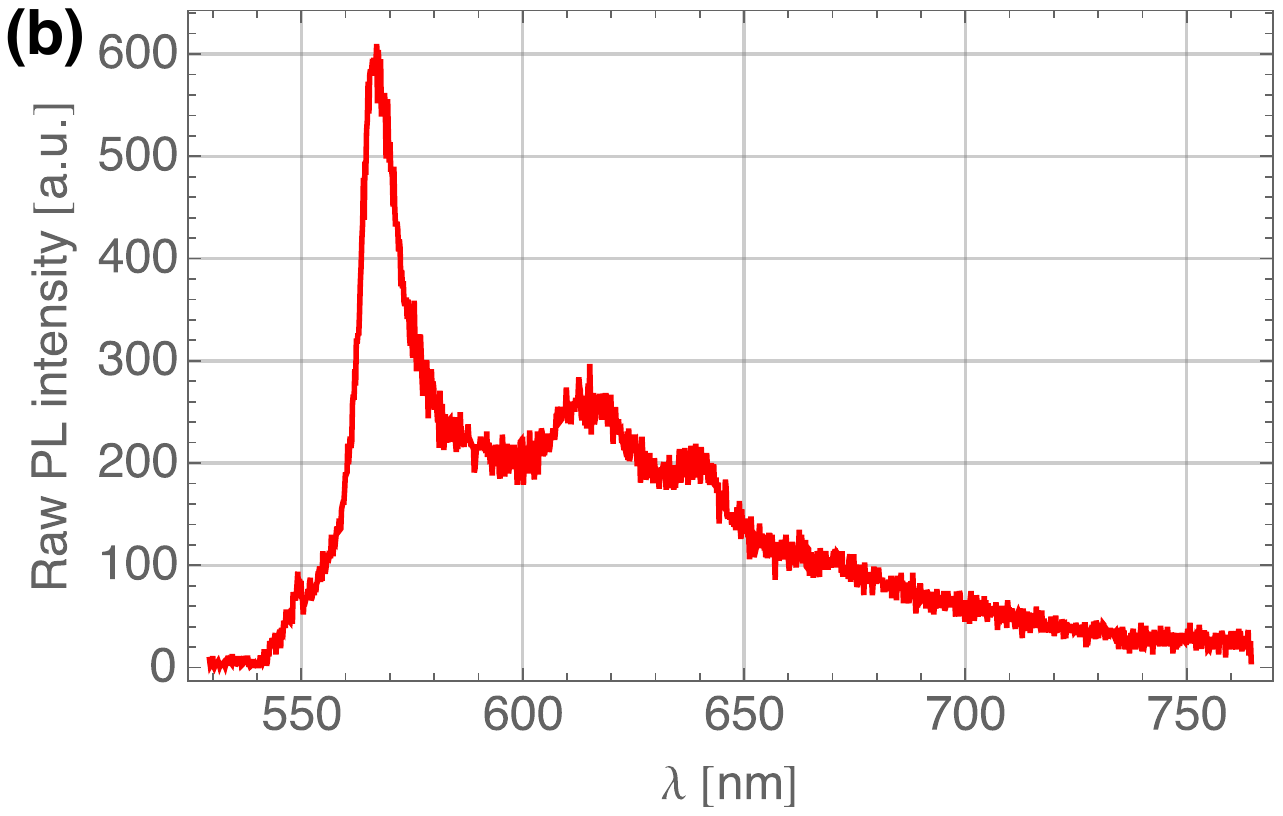}
\end{subfigure}
\begin{subfigure}{0.34\textwidth}
\centering
\includegraphics[width=0.99\linewidth,keepaspectratio]{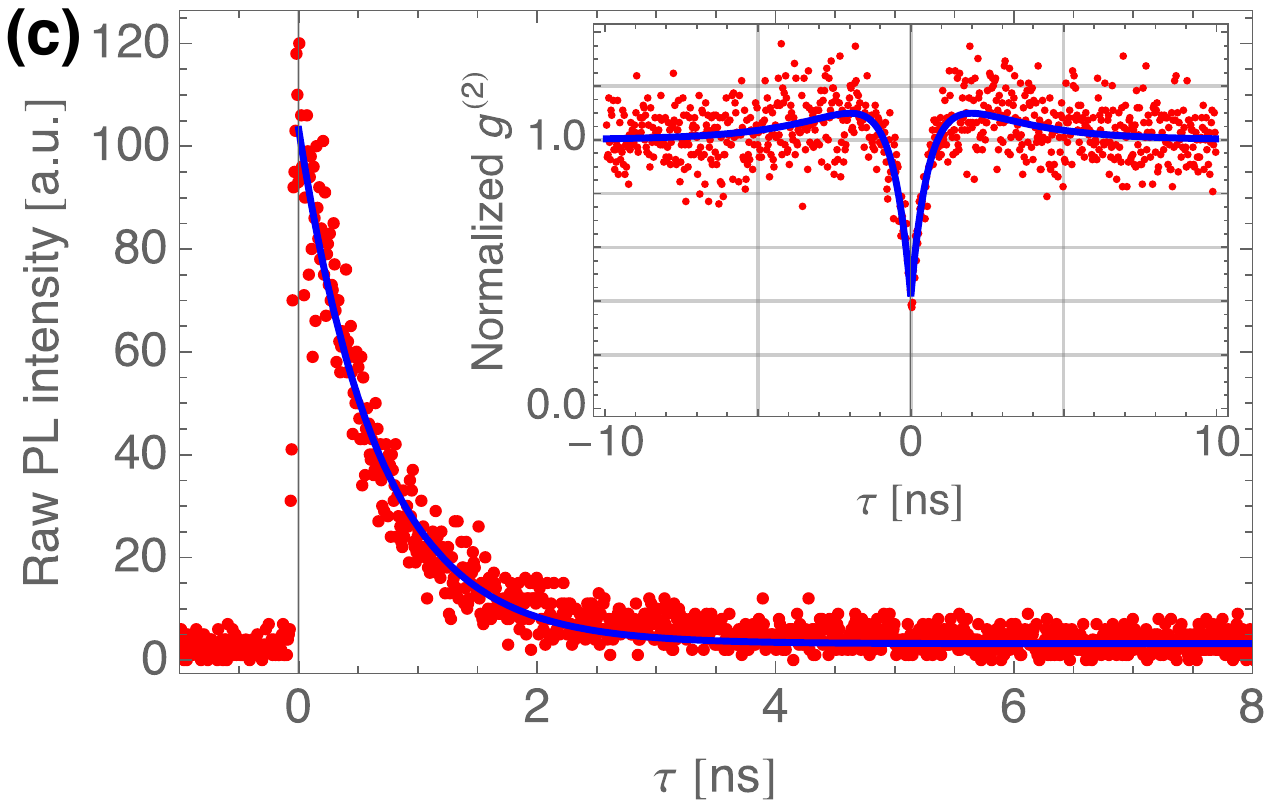}
\end{subfigure}\\
\vspace{1mm}
\begin{subfigure}{0.3\textwidth}
\centering
\includegraphics[width=0.99\linewidth,keepaspectratio]{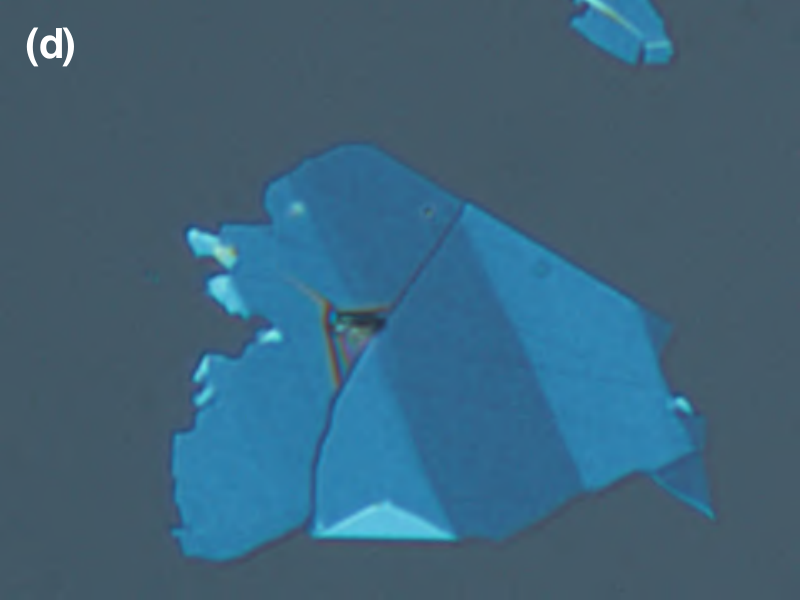}
\end{subfigure}
\begin{subfigure}{0.34\textwidth}
\centering
\includegraphics[width=0.99\linewidth,keepaspectratio]{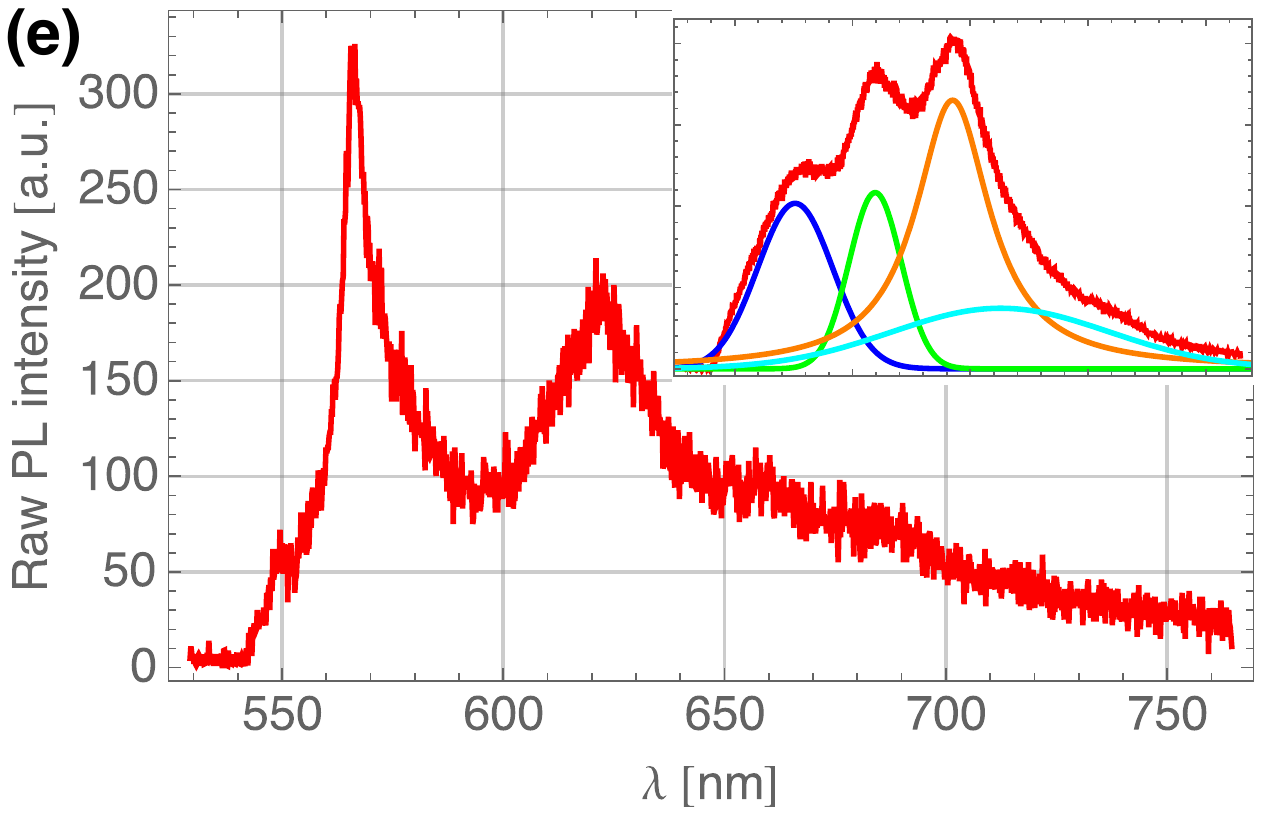}
\end{subfigure}
\begin{subfigure}{0.34\textwidth}
\centering
\includegraphics[width=0.99\linewidth,keepaspectratio]{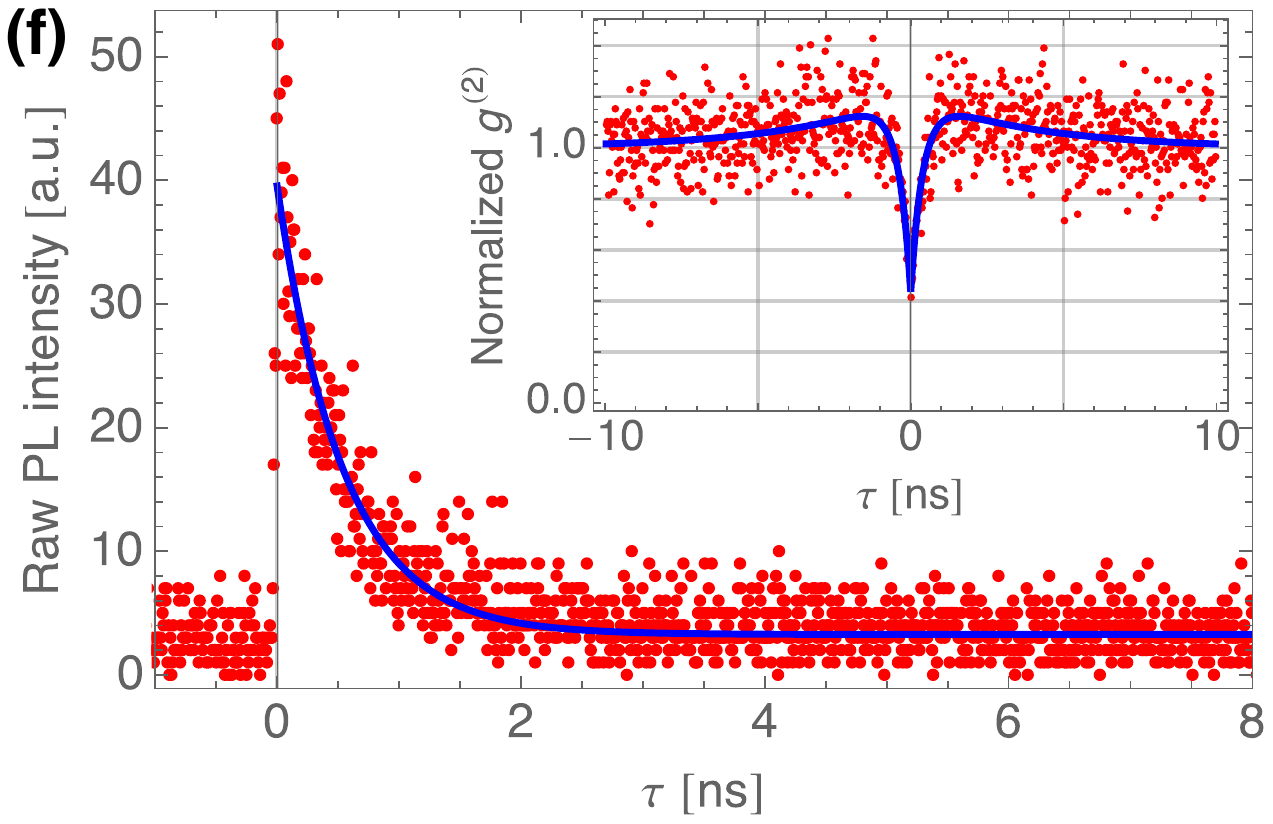}
\end{subfigure}
\caption{\textbf{Deterministic transfer of a quantum emitter. (a),(d)} Optical microscope image before and after the transfer at $1000\times$ magnification. The locations of defects D1-D2 are marked with yellow dots. \textbf{(b),(e)} Spectrum of D2 before and after the transfer. The ZPL peaks at $567.61(8)\,\text{nm}$, which is marginally blue-shifted to $567.39(13)\,\text{nm}$ after the transfer. The small inset in (e) shows the spectrum before the plasma cleaning: From a fit four peaks can be extracted, which can be assigned to the PVP (blue), NVP (green) and PVA (orange). The horizontal axis has the same scale as the large spectrum, while its vertical axis is on a much larger scale. \textbf{(c),(f)} Time-resolved photoluminescence response before and after the transfer. The excited state lifetime is $\tau=468(8)\,\text{ps}$ and is shortened to $\tau=375(15)\,\text{ps}$ after the transfer. The purity remains approximately constant (small insets), with $g^{(2)}=0.416(55)$ and $g^{(2)}=0.433(57)$ before and after the transfer, respectively.}
\label{fig:3}
\end{figure*}

\begin{figure*}[t!]
\centering
\includegraphics[width=0.75\linewidth,keepaspectratio]{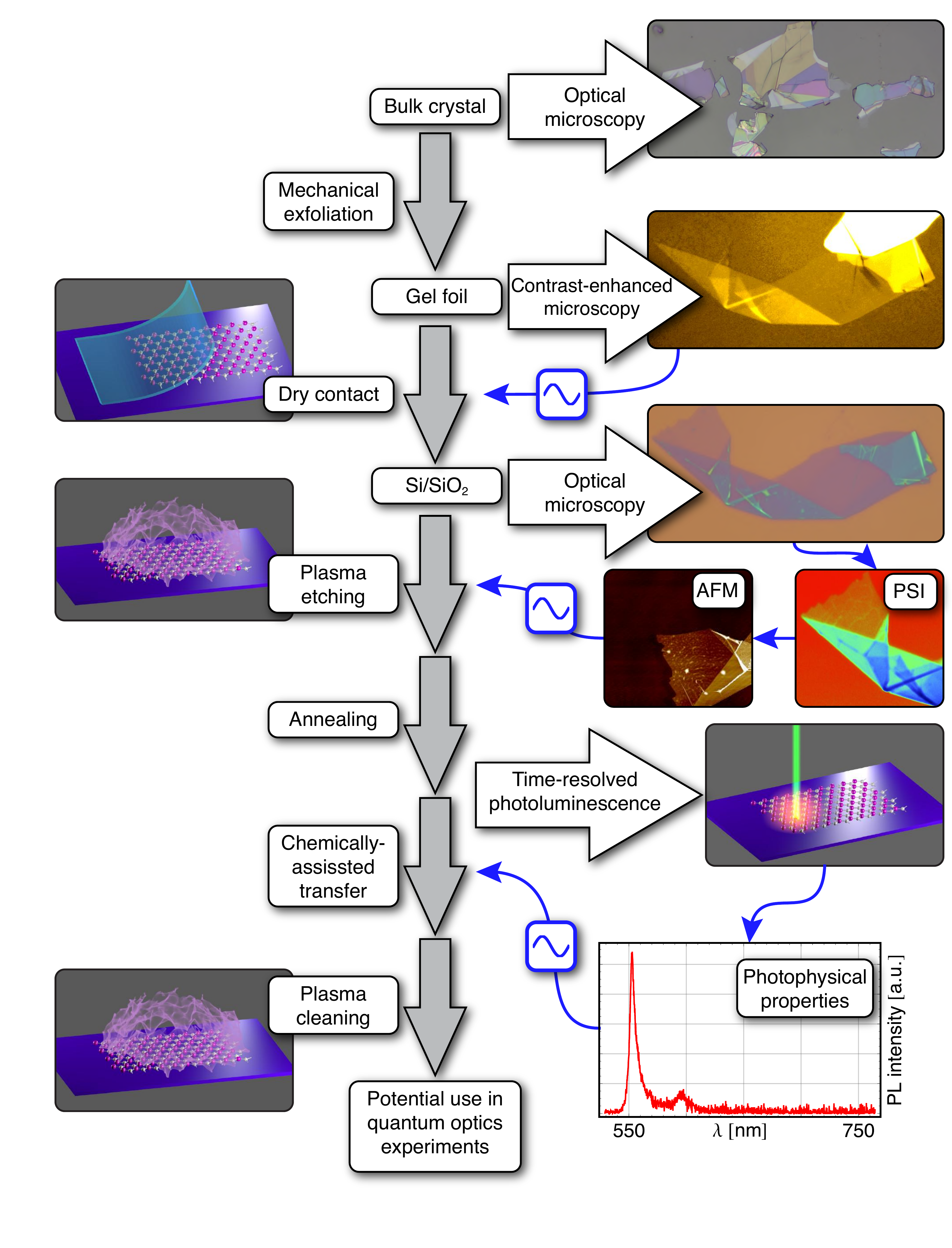}
\caption{\textbf{Full process cycle for hBN quantum emitter fabrication.} Left column shows the processes introduced and right column shows the characterization and selection of samples. The central column shows the development of the sample. hBN flakes are initially exfoliated from bulk crystal. The flakes are first optically identified using contrast-enhanced microscopy. Appropriate flakes are selected for a dry contact transfer to Si/SiO$_2$ substrates. The transferred flakes are again selected for flake thickness measurement using phase-shift interferometry (PSI). Depending on the optical path length value the exact physical thickness is measured using atomic force microscopy (AFM). Crystals with thicknesses in the suitable range undergo oxygen plasma etching and thermal annealing, after which they are fully optically characterized in a time-resolved photoluminescence (TRPL) setup. Flakes with good photophysical properties could be transferred onto waveguides or fibers, where the single photon sources could be used in a potential quantum optics experiment.}
\label{fig:4}
\end{figure*}

Finally, we demonstrate a deterministic transfer of the quantum emitters onto arbitrary substrates. The Si/SiO$_2$ substrates, on which the hBN flakes are bonded by Van der Waals force, are good for characterization, but from an application point of view, the single photon emitter must be integrated into photonic devices or networks. It is possible to transfer the flakes directly from the polymer foil onto the photonic device before plasma treatment and thermal annealing, but as the defects are formed at random positions this is not favorable. Furthermore, the high annealing temperature may damage integrated single photon devices. For monolayer transition metal dichalcogenides (TMDs) it has been demonstrated that stress induced by nanopillars allows the formation and precise positioning of quantum emitter arrays\cite{10.1038/ncomms15093}. Here we employ a wet chemical transfer method developed to transfer TMDs from SiO$_2$ onto other substrates\cite{doi:10.1021/acsnano.6b00961}. The technique is based on using a 2-component polymer mediator, which consists of polyvinylpyrrolidone (PVP) / N-vinylpyrrodoline (NVP) and polyvinyl alcohol (PVA), where the PVP/NVP provides good adhesion to the crystal while the PVA reinforces the mechanical strength of the PVP film. However, as the hBN flakes have considerably more layers compared to monolayered crystals, we adapted the polymer concentrations (see methods). The solutions are spin coated onto the sample and the resulting polymer carpet can be pressed onto a new viscoelastic foil, from which it can be transferred to its new substrate. Then the PVP/NVP dissolves easily in water.\\
\indent After the transfer all hBN crystals exhibited a strong broadband PL emission with peak maxima at $575.5\,\text{nm}$, $609.6\,\text{nm}$, $642.5\,\text{nm}$ and $662.9\,\text{nm}$ (see figure $\ref{fig:3}$(e), small inset), making it impossible to resolve the single photon ZPL. This PL was traced back to polymer chains remaining on the hBN, with PVP peaking around $576.4\,\text{nm}$, NVP peaking around $605.2\,\text{nm}$ and PVA peaking around $619.6\,\text{nm}$. The hBN red-shifts the PVA peak, explaining the third large background peak. The polymers adhered to the hBN even after soaking in distilled water for 14 hours at elevated temperatures of $60\,^{\circ}\text{C}$ for accelerated solution, meaning that the adhesion of the polymer to the hBN is stronger than its solubility in water. The solubility in other polar protic solvents (mostly alcohols) turned out to be too low as well. Finally, using low power oxygen plasma cleaning, the polymers can be fully removed. However, great care must be taken such that the hBN itself is not etched. It shall be mentioned that the 2D materials community developed a great toolbox of other transfer techniques, for example utilizing different polymers for the pick-up\cite{doi:10.1038/ncomms11894}. Using different polymers for the transfer might not introduce fluorescent residues.\\
\indent We characterized the single photon emission properties before and after a full transfer cycle. The example flake presented here hosted two defects, which both have survived the transfer. An optical microscope image prior to the transfer is shown in figure $\ref{fig:3}$(a), with both defects marked with yellow dots labeled D1-D2. The spectrum and lifetime before the transfer process are shown in figure $\ref{fig:3}$(b) and (c), respectively. The small inset in (c) is the second-order correlation function. Prior contact with the polymers the ZPL was at $567.61(8)\,\text{nm}$ with a linewidth of $4.99(17)\,\text{nm}$ and the lifetime was $468(8)\,\text{ps}$ with $g^{\left(2\right)}\left(0\right)=0.416(55)$. After the full transfer process including plasma cleaning, the microscope image shows additional cracks in the host crystal (see figure $\ref{fig:3}$(d)), but the part with the single photon emitter is not affected. Intermediate microscope images after each step show that the cracks are not caused by the polymers, but rather occur during peeling off the polymer carpet from the initial substrate. Repeated experiments proved that this happens only where the host crystal already has cracks prior to the transfer process (see figure $\ref{fig:3}$(a)). Defects which are close to such cracks are therefore not suitable for this transfer method. The ZPL is slightly blue-shifted to $567.39(13)\,\text{nm}$ with the linewidth unchanged, as shown in figure $\ref{fig:3}$(e). The defect's ZPL peak brightness is only $53.47\,\%$ of the brightness before the transfer, with the phonon sideband approximately equally strong compared to prior the transfer. This results in $13.0\,\%$ of the light being emitted into the ZPL, which was $24.9\,\%$ prior to the transfer. Narrow filtering of the ZPL nevertheless allowed measurement of the excited state lifetime, which is shortened to $375(15)\,\text{ps}$ with $g^{\left(2\right)}\left(0\right)$ increased to $0.433(57)$. The shortening of the lifetime might be due to small alterations of the host crystal structure (meaning the defect's environment) during the plasma cleaning. The same might apply for the reduction of emission into the ZPL. Further optimization of the transfer process might increase the overall performance of the transfer cycle, especially in reference to the loss in brightness of the ZPL. However, so far every transfer cycle was successful. A full process cycle starting with the bulk hBN to the chemical transfer process is shown in figure $\ref{fig:4}$.

\section*{Conclusion}
The fabrication techniques reported here demonstrate how oxygen plasma etching can create color centers in exfoliated multi-layer hexagonal boron nitride which form, after optical activation through thermal annealing, stable single photon emitter. The emitters show excellent optical properties in terms of narrow linewidths and lifetimes as short as $294\,\text{ps}$, which are one order of magnitude shorter than reported so far, allowing for a high operational bandwidth of the single photon source. Extended statistics show that many emitter with these photophysical properties are created, almost all of which have a lower time-bandwidth product at room temperature than previously reported. The emitters are also very robust, maintaining their single photon emission capabilities over the timeframe of this experimental work, which is currently 8 months. However, due to the substantial variation of even basic optical properties such as ZPL position in the spectrum or excitonic lifetime even from emitter to emitter hosted by the same flake, the exact nature of the defect remains obscure. This indicates that different defects are present, which is additionally emphasized by the fact that we did not find a correlation between single photon emission properties and the fabrication parameters or geometrical features of the host crystals. Finally, we have also demonstrated that these quantum emitters can be transferred reliably, while preserving their single photon emission capabilities. This technique allows the integration of the single photon sources into photonic circuits and networks, such as fibers and waveguide platforms. Thus, this provides a building block for next-generation quantum information processing. Only commonly available nanofrabrication processes have been used, making the fabrication easy and repeatable.\\
\newline
\textit{Note added:} While under review we became aware of a recent related work\cite{arXiv:1710.07010}.

\begin{acknowledgement}

This work was funded by the Australian Research Council (CE110001027, FL150100019 and DE140100805). We thank the ACT Node of the Australian National Fabrication Facility for access to their nano- and microfabrication facilities, particularly Kaushal Vora for technical support with the RTA and Fouad Karouta for technical support with the plasma system. We also thank Hark Hoe Tan for access to the TRPL system.

\end{acknowledgement}


\section*{Methods}
\subsection*{Sample fabrication}
The bulk crystal was acquired from HQGraphene and exfoliated to Gel-Pak WF-40-X4 and transferred  by dry contact to Si/SiO$_2$ substrates ($280\,\text{nm}$ thermally grown). After thickness measurements using PSI or AFM the samples are treated with a microwave plasmas of different powers and lengths. The pressure for all plasmas was set to $0.3\,\text{mbar}$ at an oxygen gas flow rate of $300\,\text{cm}^3/\text{min}$ at room temperature. Subsequent annealing at different temperatures for a few minutes under an Argon atmosphere takes place in a rapid thermal annealer. After multiple evacuations of any residual gases the Ar flow was set to $500\,\text{cm}^3/\text{min}$. After the annealing the samples cooled down at its natural cooling rate, without keeping the cooling rate at a maximal value.
\subsection*{Optical characterization}
The home-built micro-photoluminescence setup uses an ultrashort pulsed frequency-doubled $1044\,\text{nm}$ laser (High Q Laser URDM) focused down to the diffraction limit by an Olympus $100\times$/$\text{NA}=0.9$ dry objective. The samples are mounted on Newport piezo scanning stages with $0.2\,\mu\text{m}$ resolution. The emission, collected through the same objective, is frequency-filtered (Semrock RazorEdge ultrasteep long-pass edge filter) to dump the excitation light and guided to a spectrometer (Princeton Instruments SpectraPro). The pulse length of the laser is $300\,\text{fs}$ at a repetition rate of $20.8\,\text{MHz}$. The laser pulses are split into trigger and excitation beam and a single photon counter (Micro Photon Devices) detects the emitted photons after the trigger signal. The temporal correlation between trigger time and single photon arrival time is given by a PicoHarp 300. The second-order correlation function is measured in a different setup using a $512\,\text{nm}$ diode laser. This setup is equipped with a nanopositioning stage and a spectrometer as well. The single photon counter used in this setup are PerkinElmer SPCM-AQR-16, which is an ultralow dark count single photon counting module with dark count rates as low as $20\,\text{s}^{-1}$.
\subsection*{Transfer process}
The method was developed in $\citep{doi:10.1021/acsnano.6b00961}$, but the polymer concentrations for hBN have been adjusted. The target substrates are initially plasma cleaned. The samples are pre-baked at $80^{\circ}\text{C}$ for $1-2\,\text{min}$ and subsequently spin coated at $2000\,\text{rpm}$ for $50\,\text{s}$ with a PVP/NVP solution ($1.7\,\text{g}$ PVP + $1.5\,\text{mL}$ NVP + $0.75\,\text{mL}$ H$_2$O + $7\,\text{mL}$ Ethanol, dissolved at $40^{\circ}\text{C}$ and filtered) and then post-baked for $1-2\,\text{min}$. This is repeated with a $9\,\%$ PVA solution (molecular weight in DI water). The resulting polymer is peeled off at the edges using a scalpel and then pressed onto a new gel foil (Gel-Pak WF-40-X4) and the polymer carpet remains on the foil. Next, the crystal is transferred to a new sample by standard means. After baking the new substrate with the gel foil attached at $120^{\circ}\text{C}$ for $3-5\,\text{min}$, the polymer remains on the new substrate and is dissolved in DI water for $1\,\text{hour}$ and rinsed with IPA. A final plasma cleaning step removes remaining polymer chains.

\providecommand{\noopsort}[1]{}\providecommand{\singleletter}[1]{#1}%
\providecommand{\latin}[1]{#1}
\providecommand*\mcitethebibliography{\thebibliography}
\csname @ifundefined\endcsname{endmcitethebibliography}
  {\let\endmcitethebibliography\endthebibliography}{}


\clearpage

\end{document}